\documentclass[aps,10pt,pra,twocolumn,showpacs,floatfix,a4paper]{revtex4-1}

\usepackage{hyperref}
\usepackage{amssymb}
\usepackage{epsfig}
\usepackage{amsmath}
\usepackage{graphicx}
\usepackage{subfigure}
\usepackage{setspace}

\newcommand{\rme}{\mathrm{e}}
\newcommand{\rmd}{\mathrm{d}}
\newcommand{\rmm}{\mathrm{m}}
\newcommand{\rmc}{\mathrm{c}}
\newcommand{\rms}{\mathrm{s}}
\newcommand{\rmL}{\mathrm{L}}
\newcommand{\rmX}{\mathrm{X}}
\newcommand{\rmY}{\mathrm{Y}}

\newcommand{\rmrp}{\mathrm{rp}}
\newcommand{\rmth}{\mathrm{th}}
\newcommand{\rmpt}{\mathrm{pt}}
\newcommand{\rmcc}{\mathrm{cc}}
\newcommand{\rmin}{\mathrm{in}}
\newcommand{\rmsd}{\mathrm{sd}}
\newcommand{\rmel}{\mathrm{el}}
\newcommand{\rmdef}{\mathrm{def}}
\newcommand{\rmeff}{\mathrm{eff}}

\newcommand{\rmout}{\mathrm{out}}
\newcommand{\mcE}{\mathcal{E}}
\newcommand{\mcO}{\mathcal{O}}
\newcommand{\mcS}{\mathcal{S}}
\newcommand{\bfu}{\textbf{u}}
\newcommand{\bfE}{\textbf{E}}

\begin{document}

\title{Improving the optomechanical entanglement and cooling by photothermal force}

\author{Mehdi~Abdi}
\affiliation{Department of Physics, Sharif University of Technology, Tehran, Iran}

\author{Ali~Reza~Bahrampour}
\affiliation{Department of Physics, Sharif University of Technology, Tehran, Iran}

\begin{abstract}
Cooling and Entanglement in optomechanical systems coupled through radiation pressure and photothermal force is studied.
To develop the photothermal model, we derive an expression for deformation constant of the force.
Exploiting linearized quantum Langevin equations we investigate dynamics of such systems.
According to our analysis, in addition to separate action of radiation pressure and photothermal force, their cross correlation effect plays an important role in dynamics of the system.
We also achieve an exact relation for the phonon number of the mechanical resonator in such systems, and then we derive an analytical expression for it at weak coupling limit.
At strong coupling regime, we show that utilizing the photothermal pressure makes the ground state cooling more approachable.
The effect of photothermal force on the optomechanical entanglement is investigated in detail.
According to our exact numerical and approximate analytical studies, even though the photothermal force is naturally a dissipative force, it can improve the optomechanical entanglement both quantitatively and qualitatively.
\end{abstract}

\pacs{42.50.Wk, 03.67.Mn, 85.85.+j, 42.50.Lc}
\maketitle
%
%
\section{Introduction}
Cavity optomechanics has attracted considerable attention in the last decade.
Such a raising prominence is mainly due to its capability in providing new insight into the quantum behavior of macroscopic objects and exploring the interface between quantum and classical mechanics \cite{Kleckner2006,Groblacher2009,Groblacher2009a,Jost2009,Schliesser2008,Teufel2009}.
Alongside this fundamental quest, cavity optomechanics has many other benefits ranging from very precise measurements on forces and positions \cite{Poggio2008,Dobrindt2010,Mancini2003,Schliesser2009,Lucamarini2006} to quantum information processing \cite{Borkje2011,Pirandola2006,Vacanti2008}.
Cooling a micromechanical mirror toward its ground state, as was achieved at the beginning of this decade \cite{Teufel2011,Chan2011}, is a way of reaching the regime where quantum effects are important for a macroscopic object.
Beyond cooling a micomechanical mirror, ways of preparing such a system in a nonclassical state have also been suggested \cite{Jahne2009,Nunnenkamp2010}.
Furthermore, the entanglement between the optical field and the mechanical resonator, something which is of fundamental interest whilst also having important potential applications, has also been investigated  \cite{Vitali2007a,Genes2008,Genes2008a}.
The fast progress of nanotechnology has enabled the fabrication of high quality micromechanical resonators, which facilitates cooling of one or more natural modes of such resonators to their ground state and entangling the mechanical and optical degrees of freedom of the system.
Strong coupling between vibrational modes of a mechanical resonator to optical cavity field, which can lead to significant cooling of the mechanical mode, is demonstrated in various types of optomechanical systems \cite{Arcizet2006,Gigan2006,Metzger2004,Naik2006}.
Such a strong coupling is potentially a key feature for creating the entanglement.

From an experimental point of view, a micromechanical mirror in an optomechanical system can be cooled down by two major approaches; self cooling, introduced by Braginsky and Vyatchanin \cite{Braginsky2002}, and the cold damping.
In the latter, an artificial optical feedback is employed, which was proposed and investigated theoretically by Mancini \textit{et al} in Ref.~\cite{Mancini1998}.
Instead, in a self cooling optomechanical device, the off resonant operation of the cavity results in a retarded back action on the mechanical resonator, and hence, a self modification of its dynamics.
The optical force which affords this back action effect, couples motion of the mechanical oscillator to the cavity mode; and such a coupling can become strong in appropriate conditions.
As mentioned above, the strong optomechanical coupling leads to substantial cooling of the mechanical oscillator and the \textit{optomechanical entanglement} \cite{Mancini2002,Vitali2007a}. 
The back action is usually generated by radiation pressure and photothermal pressure, also dubbed in some literatures as bolometric force.
Depending on the geometry of the device, photothermal effects can become important, e.g. in cantilever suspended micromirrors.
For the first time it was experimentally observed in Ref.~\cite{Metzger2004} that a micromirror mounted on the cantilever of an AFM is cooled down due to photothermal effects.
An essentially similar device was used in later experiments to improve the photothermal cooling \cite{Favero2007}.

The photothermal effect was first studied theoretically via a classical approach by Metzger \textit{et al} \cite{Metzger2008}.
In addition to the cooling, the self induced oscillation, is also observed and studied as the opposite effect \cite{Metzger2008a}.
However, cooling a macroscopic or mesoscopic mechanical oscillator toward its ground state brings in quantum effects.
Therefore, it is necessary to study the problem through a quantum mechanical formalism.
An interaction Hamiltonian for the radiation pressure effect on a moving mirror is provided since 1994 \cite{Jacobs1994,Law1994,Law1995}.
These works were the basis for quantum theory of optomechanical self cooling as well as optomechanical instability in the case of radiation pressure coupling \cite{Marquardt2007,Wilson-Rae2007,Genes2008b,Ludwig2008}.
Nevertheless, a full detailed quantum mechanical theory for photothermal coupling is still lacking.
In fact, in the case of an optomechanical system with considerable photothermal effects, because of macroscopic degrees of freedom and the dissipative nature of the force, the problem is much more sophisticated.
In a semiclassical formalism, the effect was first studied by Pinard and Dantan \cite{Pinard2008}.
They introduced a phenomenological Hamiltonian for the photothermal force and analyzed the effective mechanical damping rate and effective frequency resulting from this Hamiltonian.
Starting from the same semiclassical model but via a different approach, the problem is discussed in Ref.~\cite{DeLiberato2011}.
For a setup consisting both radiative and bolometric effects, they showed that photothermal force can cool down the mechanical oscillator to only five phonons.
In an slightly modified method it was shown that one can obtain a micromechanical mirror nearly cooled to its ground state \cite{Restrepo2011}.

The strong optomechanical coupling regime where the optomechanical entanglement can be established, has not been studied in the photothermal case, so far.
In this paper, we start from the phenomenological Hamiltonian of the photothermal effect to derive the quantum Langevin equations (QLEs) of the system.
Exploiting the linearized QLEs, we elaborate a quantum mechanical analysis for photothermal optomechanics.
In fact, exact solution of the linearized QLEs provides a formalism which allows us to describe the system in strong coupling regime.
To study the optomechanical entanglement, such a formulation becomes necessary.
We shall exploit this formalism to study the photothermal cooling and the \textit{photothermal optomechanical entanglement}.
Interestingly, our results state that although photothermal process is naturally a dissipative force, yet exploiting the photothermal effect, one can increase value and robustness of the optomechanical entanglement.
Furthermore, we present a promising theoretical evidence for the photothermal cooling toward the ground state using the available devices in the present laboratories.
In fact, the photothermal force is able to significantly improve the cooling, thanks to the strong optomechanical coupling. 

The paper is organized as following:
In Sec.~II we present model of the photothermal force and introduce Hamiltonian of the system.
In Sec.~III dynamics of the system is studied through standard QLE treatment, then we calculate the correlation matrix of the system, and from position spectrum of the mechanical resonator effective damping and mechanical frequency are derived.
In Sec.~IV photothermal cooling is investigated both analytically and numerically.
The optomechanical entanglement introduced by photothermal pressure is studied in Sec.~V through numerical and analytical methods.
Finally, the paper is concluded in Sec.~VI.

%
%
\section{Model}
Essentially an optomechanical system is composed of an optical cavity associated with a mechanical degree of freedom, which up to now, it has been mounted in several different ways.
However, to deal with the problem we consider a generic and simple optomechanical model consisting a heavy fixed input mirror with a finite transparency and a light movable mirror constructing a Fabry-Perot cavity with a changeable length (see Fig.~1).
In a photothermal system, peculiarly the micromechanical mirror is not perfectly reflective, so it absorbs the intracavity photons.
The absorption process leads to excitation of atomic structure of the mirror, resulting an increment in the number of free electrons, which instead increases temperature of the mirror.
In fact, these photons are mostly absorbed at its surface as skin depth of the mirror is typically very small compared to the dimension of the mirror.
The heat diffuses from the surface into the mirror through a conduction process.
Therefore, a temperature gradient develops inside the mirror, which is a source for elastic waves (called thermoelastic waves) \cite{Landau1970}.
In return, the elastic waves modify length of the optical cavity leading to change of the cavity resonance frequency.
Therefore, the position of the mirror reacts on the cavity field amplitude.
The deviation of the micromechanical mirror from its equilibrium position is a superposition of photothermal modifications, $q_{\rmpt}$, and radiation pressure deformations, $q_{\rmrp}$ [see Fig.~1]. 

It is worth mentioning that photothermal effects only become important in optomechanical devices in which there is significant absorption of the intracavity photons by the mechanical oscillator.
Strictly speaking, in every optomechanical system both photothermal and radiation pressure back actions contribute in dynamics of the system.
However, the amount of their contribution is different in various setups.
It is practically impossible --or at least very difficult-- to prepare a setup merely based on the radiation pressure or the photothermal pressure.
Nevertheless, in some schemes the photothermal effects are negligible, e.g. an optical cavity with a dielectric membrane in the middle \cite{Thompson2008}, suffers small photothermal effects.

\begin{figure}
\label{scheme}
\includegraphics[width=3.4in]{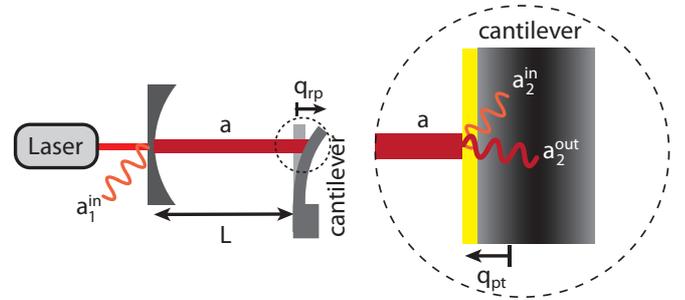}
\caption{(Color online) Scheme of a cantilever suffering photothermal effects: the radiation pressure affects on the position of the cantilever due to its flexibility, while the photothermal force changes the cavity length by the surface thermoelastic waves.}
\end{figure}

\subsection{Photothermal force}
It is convenient to consider the photothermal problem effectively as a feedback force on the movable mirror.
In fact, it is very difficult to introduce a full quantum mechanical model for the photothermal process as dealing with it needs to consider a large number of degrees of freedom; phonons and electrons, and their interaction.
Therefore, we use a semiclassical model for it.
As mentioned above, the effect is a complicated process starting from absorption of intracavity photons, leading to modification of the cavity field amplitude.
Obviously, amplitude of the cavity field responds to these processes with a time lag.
Such a delay is dominantly determined by thermal properties of the absorbing material.
These processes can be encapsulated in a simple formalism describing the action of absorbed photons on the mechanical oscillator with an appropriate response function \cite{Metzger2008}.
Therefore, the feedback model of the bolometric force is written in the following form
\begin{equation}
F_{\rmpt}(q(t))=\int_{-\infty}^{\infty}{g(t-s) \frac{\rmd F_{\rmdef}(q,s)}{\rmd s}\rmd s},
\label{PT}
\end{equation}
where $g(t) \equiv \Theta(t)(1-\exp\{-t/\tau_{\rmth}\})$ is the response function of the system with a thermal characteristic time $\tau_{\rmth}$, determined by the heat diffusion procedure.
The instantaneous deformation force $F_{\rmdef}$ in Eq.~(\ref{PT}) is phenomenologically proportional to the absorbed optical energy.
Therefore, the rate of photon absorption --having in mind that the absorption process in the mirror takes place with a quantum efficiency equal to $\beta$-- is $\beta I_{2}^{\rmout}(t)$ with $I_{2}^{\rmout}(t)=a_{2}^{\rmout,\dagger}(t)a_{2}^{\rmout}(t)$, where $a_{2}^{\rmout}=\sqrt{2\gamma_2}~a - a_{2}^{\rmin}$ is the output field at the micromechanical mirror side \cite{Walls2008}.
Taking into account adjustment of the intracavity photon energy by position of the mirror, the deformation force reads
\begin{equation}
F_{\rmdef}(q,t)=\chi \hbar \omega_{\rmc}(q) \beta I_{2}^{\rmout}(t)
\approx \chi \beta \hbar \omega_{\rmc} [1-\frac{q(t)}{L}]I_{2}^{\rmout}(t),
\label{deformforce}
\end{equation}
where, $\chi$ is the constant of proportionality and its dimension is inverse of velocity.
The deformation constant $\chi$ is discussed in the next subsection where an explicit expression for it is derived.

Integrating by parts, the photothermal force turns from Eq.~(\ref{PT}) to
\begin{align}
\label{photoforce}
F_{\rmpt}(q(t)) &=F_0 
-\int_{-\infty}^{\infty}{\frac{\rmd g(t-s)}{\rmd s}F_{\rmdef}(q,s)\rmd s} \\
&=F_0 +\frac{1}{\tau_{\rmth}} \int_{-\infty}^{\infty}{\Theta(t-s)
 \rme^{-\frac{t-s}{\tau_{\rmth}}} F_{\rmdef}(q,s)\rmd s}, \nonumber
\end{align}
where $F_0$ is a constant force whose only effect is modification of the mechanical equilibrium position.

\subsection{Deformation constant}
Let us discuss about the deformation constant introduced in the photothermal force equation (\ref{deformforce}).
In Ref.~\cite{Pinard2008} through a phenomenological assumption the constant was taken equal to $2/c$, where $c$ is the speed of light.
Instead, in Ref.~\cite{Restrepo2011}, the authors argued that the constant considered in Ref.~\cite{Pinard2008} does not provide the actual strength of the force.
Therefore, they introduced a factor which pertains to the quantum efficiency of the photon absorption (in our notation it is $\beta$) yet it can become much greater than unity.
On the other hand, De Liberto \textit{et al} obtained a value for $\chi$ by fitting their theory to the experimental data \cite{DeLiberato2011}.
Here, we shall suppose that $\beta$ gets values between 0 and 1, and then we show that $\chi$ essentially relates to elasticity of the micromechanical mirror.
Starting from the thermoelastic wave equation, we find an explicit expression for $\chi$.
In the following lines, we temporarily suppress radiation pressure effects and focus only on the photothermal force.

The three dimensional thermoelastic wave equation can be split into longitudinal and transversal components respect to its direction of propagation (see appendix~A).
Here we are only concerned about the longitudinal component, since only it is responsible for modifications of the cavity length.
Therefore, assuming $x$ as symmetry axis of the cavity, the thermoelastic wave equation is given by
\begin{equation}
\ddot{u}_{x} -c_{x}^{2} \frac{\partial^{2} u_{x}}{\partial x^{2}} =-\frac{E_{\rmY} \alpha_{\rmth}}{3 \varrho(1-2\sigma)} \frac{\partial T}{\partial x},
\label{longwave}
\end{equation}
where $E_{\rmY}$ is Young modulus, $\sigma$ is Poisson ratio, $\alpha_{\rmth}$ is linear heat expansion coefficient of the micromechanical mirror and $\varrho$ is its density.
Note that the source term on the right hand side of Eq.~(\ref{longwave}) involves the temperature gradient inside the mirror which in turn stems from the heat diffusion.
For the sake of simplicity, we have assumed that there is a temperature gradient only along the cavity axis.
This is reasonable for a uniform heat flux on the surface of the mirror.
Since surface of the mirror is heated by the photon absorption process, considering a uniform cavity field profile makes our supposition reliable.
In fact, one can define an effective radius for the cavity beam profile (which is typically a Gaussian) which its prevailing power appears inside a circle with radius $r_{0}$.
Thus, this assumption is valid for a mirror with area small compared to cross sectional area of the cavity beam.

On the other hand, neglecting effect of deformations on the temperature of the material, thermal diffusion equation is given by
\begin{equation}
\varrho C \frac{\partial T}{\partial t} = K_{\rmth} \frac{\partial^2 T}{\partial x^{2}} +Q_{\rmth},
\label{diffusion}
\end{equation}
where $C$ is specific heat capacity of the material, $K_{\rmth}$ is its thermal conductivity and $Q_{\rmth}$ is the heat flux.
The heat flux is given by
\begin{equation}
Q_{\rmth}(x,t)=\beta \gamma_2 n_0 \epsilon_0 |\bfE(t)|^{2} \delta_{\rmsd} \delta[x-q(t)],
\label{heatflux}
\end{equation}
where $n_0$ is the refraction index and $\delta_{\rmsd}=(2 / \mu_0 \sigma_{\rmel} \omega_{\rmL})^{1/2}$ is the skin depth of the material for frequency of the laser $\omega_{\rmL}$.
Also, $\sigma_{\rmel}$ is electrical conductivity of the mirror, and $\bfE(t)$ is the classical intracavity electric field.
In photothermal setups the surface of the cantilever is usually coated by a thin layer of gold (see e.g. \cite{Favero2007}) which its penetration depth for optical frequencies is about a few nanometers.
Hence, it is sensible to assume that the cavity photons are absorbed just on the surface of the mirror.
This assumption is imposed through the Dirac delta function in Eq.~(\ref{heatflux}).

From (\ref{diffusion}) and (\ref{heatflux}) an expression for the temperature gradient can be derived, and then replaced in the right hand side of the elastic wave equation (\ref{longwave}).
To perform Eq.~(\ref{longwave}) we use spatial Fourier transform of the longitudinal deformation $u_{x}$, defined as
\begin{equation}
u_{x}(x,t) =\int \rmd k_{x} \rme^{-i k_{x} x} \bar{u}_{x} (k_{x},t).
\end{equation}
It is clear that the cavity length is only modified by amplitude of the elastic waves on the surface of the mirror, therefore, one sets $x=0$ to arrive at
\begin{equation}\label{integ}
\int{\rmd k_{x} \Big\{\ddot{\bar{u}}_{x} +\omega_{x}^2 \bar{u}_{x} -\frac{\beta \gamma_{2} E_{\rmY} \alpha_{\rmth} \delta_{\rmsd}}{3 \varrho K_{\rmth} (1-2\sigma)} n_0 \epsilon_0 |\bfE(t)|^2} \Big\} =0,
\end{equation}
where we have introduced the longitudinal elastic wave frequency as $\omega_{x} \equiv k_{x} c_{x}$.
Now let us define the position of the micromechanical mirror $q(t)$ as superposition of elastic wave components on the surface, that is
\begin{equation}
q(t) \equiv \int \bar{u}_{x}(k_{x},t) \rmd k_{x}.
\end{equation}
Therefore, including the fluctuation and dissipation processes, from Eq.~(\ref{integ}) one arrives at the following equation of motion
\begin{align}
\ddot{q} + \gamma_{x} \dot{q} +\omega_{x}^{2} q = \frac{1}{m} (\mathcal{F}_{\rmth} +\mathcal{F}_{\rmpt}),
\end{align}
where $\mathcal{F}_{\rmth}$ is a classical thermal Langevin force and
\begin{equation}
\mathcal{F}_{\rmpt}=\frac{\pi r_{0}^{2} l_{\rmth} \beta E_{\rmY} \alpha_{\rmth} \delta_{\rmsd}}{3 K_{\rmth} (1-2\sigma)}\gamma_2 n_0 \epsilon_0 |\bfE(t)|^2,
\label{classphoto}
\end{equation}
is a classical interpretation for the photothermal force, whose time lag is implied in $|\bfE(t)|^2$ (see Ref.~\cite{Bahrampour2011}).
A comparison between (\ref{classphoto}) and (\ref{deformforce}) leads us to the following relation for the deformation constant
\begin{equation}\label{chi}
\chi = \frac{E_{\rmY} \alpha_{\rmth} \delta_{\rmsd}}{3 K_{\rmth} (1-2\sigma)} ~.
\end{equation}

We consider the photothermal setup in Ref.~\cite{Favero2007} to estimate the value of $\chi$.
The cantilever utilized in Ref.~\cite{Favero2007} is composed of a silicon substrate with a few nanometer gold coating.
This gold layer is too thin to contribute significantly in elastic vibrations of the cantilever.
Actually, the temperature gradient effectively ranges from the surface of the mirror through the thermal diffusion length $l_{\rmth} =(2 \pi K_{\rmth}/\varrho C \omega_{\rmL})^{1/2}$.
One can easily estimate that the thermal diffusion length of gold is a few micrometers, while as mentioned above the gold coating on the cantilever is just a few nanometers.
Thus, the silicon substrate dominantly contributes in the thermal expansion (and contraction) of the mirror.
Inserting thermal and elastic parameter values of silicon into Eq.~(\ref{chi}) one gets $\chi \sim 10^{-5}$, which is almost the value obtained by the curve fitting of Ref.~\cite{DeLiberato2011} and considered phenomenologically in Ref.~\cite{Restrepo2011}.

\subsection{Hamiltonian of the system}
As discussed above, in an optomechanical system the effect of the bolometric force can be realized as a feedback process.
Therefore, the effective photothermal Hamiltonian can be written from Eq.~(\ref{photoforce}) as
\begin{equation}
\label{photohamilton}
H_{\rmpt}(t) =-\frac{\hbar \omega_{\rmc}\chi \beta}{\tau_{\rmth}} q(t) \int{h(t-s)\big[1-\frac{q(s)}{L}\big]I_{2}^{\rmout}(s)\rmd s},
\end{equation}
where $\omega_{\rmc} \chi \beta/\tau_{\rmth}$ functions as the feedback loop strength and $I_{2}^{\rmout}(s)$ as the corresponding output photocurrent.
In practice, this photocurrent is absorbed by the mirror and acts on momentum of the mechanical resonator.
Moreover, the response function of the system from (\ref{photoforce}) is $h(t)=\Theta(t) \exp\{-t/\tau_{\rmth} \}$.
To investigate quantum dynamics of the system, we must write the Hamiltonian of the whole system.
We consider an optomechanical system composed of a mechanical resonator with frequency $\omega_{\rmm}$ which is subject to radiation pressure and photothermal pressure due to an optical cavity mode with frequency $\omega_{\rmc}$ which is driven by an intense laser.
It is convenient to transform canonical operators of the mechanical oscillator so that $[q,p]=i$.
Hence, the Hamiltonian reads
\begin{align}
\label{hamiltonian}
H =&~\hbar \omega_{\rmc} a^{\dagger}a+\frac{1}{2}\hbar \omega_{\rmm}(p^2+q^2) 
-\hbar G_0 a^{\dagger} a q \nonumber \\
&-\hbar \lambda q \int{h(t-s) \Big[1-\frac{G_0 q(s)}{\omega_{\rmc}}\Big] I_{2}^{\rmout}(s)\rmd s} \nonumber \\
&+i \hbar \mcE (a^{\dagger} \rme^{-i \omega_{\rmL} t} -a \rme^{i \omega_{\rmL} t}),
\end{align}
where the first term describes the cavity mode, with annihilation operator $a$ ($[a,a^{\dagger}]=1$), while the second term corresponds to the free mechanical oscillator.
Third term of (\ref{hamiltonian}) is the radiation pressure interaction, while the fourth represents the photothermal Hamiltonian.
The last one is the effective Hamiltonian of a input laser with power $P$ and frequency $\omega_{\rmL}$.
In Eq.~(\ref{hamiltonian}), $G_0 \equiv (\omega_{\rmc}/L)\sqrt{\hbar/m \omega_{\rmm}}$ is the single photon radiation pressure coupling and $\lambda \equiv (\chi \beta \omega_{\rmc}/\tau_{\rmth})\sqrt{\hbar/m \omega_{\rmm}}=\chi \beta L G_{0}/\tau_{\rmth}$ is the photothermal coupling strength, while the mean amplitude of the input field is related to the laser power by $\mcE=\sqrt{2\gamma_{1}P/\hbar\omega_{\rmL}}$, where $\gamma_{1}$ is the cavity loss rate through its input port.
We notice that since typical optomechanical parameters suggest that $G_0 q \ll \omega_{\rmc}$, we can neglect the second term of integrand in the photothermal Hamiltonian.

%
%
\section{System Dynamics}
\subsection{Quantum Langevin equations}
A full description for dynamics of the optomechanical system is provided by its QLEs which can be derived from Hamiltonian (\ref{hamiltonian}).
Fluctuation and dissipation processes can be taken into account in a fully consistent way \cite{Giovannetti2001}.
We write these QLEs in rotating frame of the laser frequency as the following
\begin{subequations}
\label{nonlinear}
\begin{eqnarray}
\dot{q}&=&\omega_{\rmm}p, \\
\dot{p}&=&-\omega_{\rmm}q -\gamma_{\rmm}p +G_0 a^{\dagger}a \nonumber \\
&&+\frac{\chi \beta L}{\tau_{\rmth}}G_{0} \int{h(t-s) I_{2}^{\rmout}(s)\rmd s} +\xi, \\
\dot{a}&=&-(\gamma_{\rmc} +i \Delta_0) a +i G_0 q a + \mcE +\sqrt{2\gamma_1} ~a_{1}^{\rmin} \nonumber\\ 
&&+\sqrt{2\gamma_2} ~a_{2}^{\rmin} ,
\end{eqnarray}
\end{subequations}
where $\Delta_0 \equiv \omega_{\rmc} -\omega_{\rmL}$ is detuning of the laser frequency from the corresponding cavity mode, while $\gamma_{1}$ and $\gamma_{2}$ are the cavity decay rates from the heavy input mirror and the micromechanical mirror, respectively, constructing the entire cavity decay rate $\gamma_{\rmc}=\gamma_1 +\gamma_2$.
Furthermore, $\xi$, $a_{1}^{\rmin}$, and $a_{2}^{\rmin}$ are the zero mean value Langevin noise operators affecting the system:
The mechanical mode is affected by a Brownian stochastic force $\xi(t)$, obeying the correlation function \cite{Gardiner2000}
\begin{equation}
\langle\xi(t)\xi(t')\rangle = \frac{\gamma_{\rmm}}{\omega_{\rmm}}
\int\frac{\omega\rmd\omega}{2\pi} \rme^{-i\omega(t-t')}\big[1+\coth(\frac{\hbar\omega}{2k_{\mathrm{B}}T})\big],
\end{equation}
where $k_{\mathrm{B}}$ is the Boltzmann constant and $T$ is the temperature of the mechanical reservoir.
The cavity field vacuum input noises have the following nonzero correlation functions
\begin{subequations}
\label{cavity}
\begin{eqnarray}
\langle a_{k}^{\rmin}(t) a_{k}^{\rmin,\dagger}(t')\rangle &=&[\bar{N}(\omega_{\rmc})+1]\delta(t-t'), \\
\langle a_{k}^{\rmin,\dagger}(t) a_{k}^{\rmin}(t')\rangle &=&\bar{N}(\omega_{\rmc})\delta(t-t'),
\end{eqnarray}
\end{subequations}
with $k=1,2$ assigning the two noise ports [see Fig.~1], and $\bar{N}(\omega_{\rmc}) = [\exp\{\hbar\omega_{\rmc}/k_{\mathrm{B}}T\} -1]^{-1}$ is the mean intracavity photon number at thermal equilibrium with temperature $T$.
At optical frequencies $\hbar\omega_{\rmc}/k_{\mathrm{B}}T \gg 1$ and therefore $\bar{N}(\omega_{\rmc}) \approx 0$, so that only the correlation function of Eq.~(\ref{cavity}a) is relevant.

As mentioned above, our aim is to focus on the strong optomechanical regime  (both radiatively and photothermally) which establishes strong quantum correlations between the steady state mechanical and optical fluctuations.
It is shown in Refs.~\cite{Vitali2007a} and \cite{Genes2008} that this is attained for very intense intracavity fields, i.e., for high finesse cavities and enough driving powers.
In this case, when the system is stable, it reaches a steady state which is characterized by the cavity mode in a coherent state, and the mechanical resonator at a new equilibrium position which is determined by the stationary intracavity photon number.
Thus, it is enough to focus onto the dynamics of the quantum fluctuations around their classical steady state.
This situation simplifies the exact QLEs in (\ref{nonlinear}) to a set of linearized equations for quantum fluctuations of the system.
Therefore, one replaces a generic operator $\mcO$ with $\mcO=\mcO_{\rm s}+\delta \mcO$ in Eq.~(\ref{nonlinear}), and gets the steady state values as $\alpha_{\rms}=\mcE/(\gamma_{\rmc} +i \Delta)$ and $q_{\rms}=\alpha_{\rms}^2 G_0(1+2\gamma_2 \chi\beta L)/\omega_{\rmm}$.
Here the effective cavity detuning is introduced as $\Delta \equiv \Delta_0 -G_0 q_{\rms}$, meanwhile we have chosen the phase of $\mcE$ such that $\alpha_{\rms}$ is real and positive without loss of generality.
Thus, dynamics of the quantum fluctuations are given by the following equations 
\begin{subequations}\label{linear}
\begin{eqnarray}
\delta \dot{q} &=& \omega_{\rmm} \delta p, \\
\delta \dot{p} &=& \frac{\chi \beta L}{\tau_{\rmth}}G\int{h(t-s)\{2\gamma_2 \delta x
-\sqrt{2 \gamma_2} x_{2}^{\rmin}\}\rmd s} \nonumber \\
&&-\omega_{\rmm} \delta q - \gamma_{\rmm} \delta p + G \delta x +\xi, \\
\delta \dot{x} &=& -\gamma_{\rmc} \delta x +\Delta \delta y +\sqrt{2 \gamma_1} x_{1}^{\rmin}
+\sqrt{2 \gamma_2} x_{2}^{\rmin}, \\
\delta \dot{y} &=& -\gamma_{\rmc} \delta y -\Delta \delta x + G \delta q +\sqrt{2 \gamma_1} y_{1}^{\rmin} +\sqrt{2 \gamma_2} y_{2}^{\rmin},
\end{eqnarray}
\end{subequations}
where the effective optomechanical coupling is defined as $G \equiv \sqrt{2} ~\alpha_{\rms} G_0$.
In fact, $G$ is a measure for optomechanical coupling strength which is dominantly determined by number of intracavity photons.
Furthermore, in (\ref{linear}) we have introduced the cavity field quadratures $\delta x \equiv (\delta a +\delta a^{\dagger})/\sqrt{2}$ and $\delta y \equiv (\delta a -\delta a^{\dagger})/i \sqrt{2}$ and corresponding noise operators $x_{k}^{\rmin}$, $y_{k}^{\rmin}$ with $k=1,2$ .

\subsection{Effective damping and optical spring effect}
Since the dynamics of the system is linearized and also the system noises are Gaussian, we can focus on a set of states; that is Gaussian states which are very interesting both experimentally and theoretically.
In particular, the bipartite steady state of the system formed by the mechanical oscillator and the cavity mode is a zero mean Gaussian state \cite{Vitali2007a}.
Hence, this state is fully characterized by its $4\times 4$ correlation matrix (CM).
Defining vector of the system operators as $u(t)\equiv[\delta q(t), \delta p(t), \delta x(t), \delta y(t)]^{\mathsf{T}}$, the CM is given by
\begin{equation}
\label{cm}
V_{ij}=\frac{1}{2}\big\langle u_i(\infty) u_j(\infty) +u_j(\infty) u_i(\infty) \big\rangle.
\end{equation}
One may determine the CM of the system by solving Fourier transform of Eqs.~(\ref{linear}), which are the following equations
\begin{subequations}
\label{frequency}
\begin{eqnarray}
-i \omega \delta \tilde{q} &=& \omega_{\rmm} \delta \tilde{p}, \\
-i \omega \delta \tilde{p} &=& -\omega_{\rmm} \delta \tilde{q} -\gamma_{\rmm} \delta \tilde{p} 
 +G \delta \tilde{x} \nonumber \\
 &&+\sqrt{2} \alpha_{\rms} \lambda \tilde{h}(\omega) \big( 
 2 \gamma_2 \delta \tilde{x} -\sqrt{2\gamma_2} \tilde{x}_{2}^{\rmin} \big) +\tilde{\xi}, \\
-i \omega \delta \tilde{x} &=& -\gamma_{\rmc} \delta \tilde{x} +\Delta \delta \tilde{y} +\sqrt{2\gamma_1} \tilde{x}_{1}^{\rmin} +\sqrt{2\gamma_2} \tilde{x}_{2}^{\rmin}, \\
-i \omega \delta \tilde{y} &=& -\gamma_{\rmc} \delta \tilde{y} -\Delta \delta \tilde{x} +G \delta \tilde{q} +\sqrt{2\gamma_1} \tilde{y}_{1}^{\rmin} +\sqrt{2\gamma_2} \tilde{y}_{2}^{\rmin}.
\end{eqnarray}
\end{subequations}
Subsequently, the CM elements can be obtained as
\begin{equation}
\label{fcm}
V_{ij}(t)=\iint{\frac{\rmd \omega \rmd \omega'}{4 \pi} \rme^{-i(\omega+\omega')t}\big\langle \tilde{u}_i(\omega) \tilde{u}_j(\omega') +\tilde{u}_j(\omega') \tilde{u}_i(\omega) \big\rangle},
\end{equation}

In particular, the first element of the CM which corresponds to covariance of the mechanical position, $\langle \delta q^2 \rangle$, reads
\begin{equation}
\label{varq}
\langle \delta q^2 \rangle = \int{\frac{\rmd \omega}{2 \pi} \mcS_{qq}(\omega) },
\end{equation}
where $\mcS_{qq}(\omega)$ is position spectrum of the mechanical resonator, given by
\begin{equation}
\label{sqq}
\mcS_{qq}(\omega) = |\rmX_{\rmeff}(\omega)|^2 \big[\mcS_{\rmth}(\omega)+\mcS_{\rmrp}(\omega)+\mcS_{\rmpt}(\omega)+\mcS_{\rmcc}(\omega)\big],
\end{equation}
where $\rmX_{\rmeff}(\omega)$ is the effective mechanical susceptibility.
The spectrum of thermal Brownian force is \cite{Gardiner2000,Landau1958}
\begin{equation}
\label{sth}
\mcS_{\rmth}(\omega)= \frac{\gamma_{\rmm} \omega}{\omega_{\rmm}}  \coth[\frac{\hbar \omega}{2 k_{\mathrm{B}} T}],
\end{equation}
while the radiation pressure and the photothermal spectra are given respectively by the following
\begin{subequations}\label{srppt}
\begin{eqnarray}
\mcS_{\rmrp}(\omega) &=& \frac{\gamma_{\rmc} G^2 (\gamma_{\rmc}^2+\Delta^2+\omega^2)}{[\gamma_{\rmc}^2+(\omega-\Delta)^2][\gamma_{\rmc}^2+(\omega+\Delta)^2]}, \\
\mcS_{\rmpt}(\omega) &=& \frac{\gamma_{2}G^{2} (\chi \beta L)^{2}}{1+\tau_{\rmth}^2 \omega^2}
\end{eqnarray}
\end{subequations}
Finally, cross correlation of the radiation pressure and photothermal is introduced by the following spectrum
\begin{align}\label{scc}
\mcS_{\rmcc}(\omega) =& 2 \gamma_2 G^{2} \chi \beta L \\
& \times \frac{\gamma_{\rmc}(\gamma_{\rmc}^2 +\Delta^{2} +\omega^2) -\tau_{\rmth} \omega^2(\gamma_{\rmc}^2 -\Delta^{2} +\omega^2)}{(1+\tau_{\rmth}^2 \omega^2)[\gamma_{\rmc}^2+(\omega-\Delta)^2][\gamma_{\rmc}^2+(\omega+\Delta)^2]}. \nonumber
\end{align}
In fact, the photothermal force in an optomechanical system --in addition to its exclusive effect-- interacts with the radiation pressure.
Such a cross correlation effect means that these two light forces do not act on the mechanical resonator, independently.
This is not surprising, since from (\ref{linear}) one can find out that both radiative and photothermal forces are coupled to the mechanical resonator through in phase quadrature of the cavity field $\delta x$ .
However, comparing Eqs.~(\ref{srppt}b) and (\ref{scc}), the importance of the joint effect may be clarified.
By investigation, one finds out that around the natural frequency of the mechanical resonator $\omega_{\rmm}$ the crucial factor is $\beta\chi L$ whose quantity for typical photothermal setups is much less than unity.
Thus, at least around mechanical frequency the cross correlation effect has more contribution than the pure photothermal spectrum.
Nonetheless, around the mechanical frequency, the cross correlation spectrum is very sensitive to $\gamma_{1}$, which limits the value of the cross correlation spectrum.
To visually show this, we compare the three spectra in Fig.~2, from which it is obvious that especially around the mechanical frequency, the cross correlation spectrum is even larger than the photothermal spectrum.
One can also see from the figure that destroying the total noise spectrum at high frequencies is the other consequence of the cross correlation effect.
In Fig.~2 we have considered a set of parameters which are generally available in laboratories (see caption of Fig.~2).

\begin{figure}
\includegraphics[width=3.1 in]{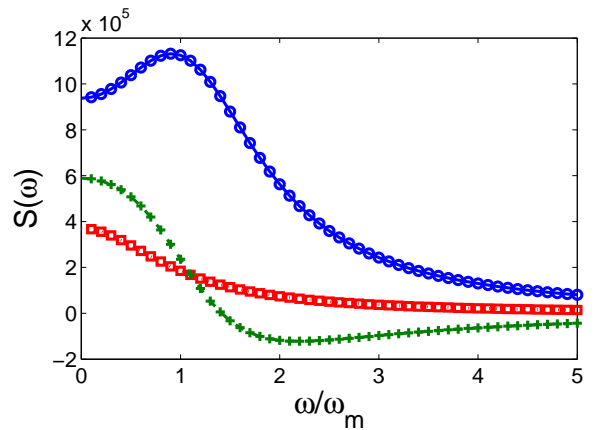}
\caption{(Color online) Mechanical Spectrum composition; radiation pressure (blue circle), photothermal force (red square), cross correlation effect (green plus). The system is composed of a mechanical oscillator with natural frequency $\omega_{\rmm}/2\pi=10$ MHz, quality factor $Q_{\rmm}=10^5$, and mass $m=5$ ng. Temperature of the reservoir interacting with the mechanical resonator is $T=0.4$ K. Deformation constant of the cantilever is $\chi=1 \times 10^{-5}$ s/m, while its quantum efficiency is $\beta=1$ and the thermal diffusion time $\tau_{\rmth} \omega_{\rmm}=1$. The cavity length is $L=1$ mm and it is pumped by a laser with wavelength $\lambda_{\rmL}=810$ nm and power of $P=1$ mW. The laser detuning from the corresponding cavity mode is $\Delta/\omega_{\rmm}=1$. Finally, the optical damping rates are $\gamma_1/\omega_{\rmm}=0.5$ and $\gamma_2/\gamma_1=1$ .}
\end{figure}

The effective mechanical susceptibility introduced in Eq.~(\ref{sqq}) is defined as
\begin{align}
\label{suscept}
\rmX_{\rmeff}(\omega) \equiv &\frac{1}{\omega_{\rmm}}\Big[\Omega_{\rmeff}^2 -\omega^2 -i \omega \Gamma_{\rmeff}\Big]^{-1},
\end{align}
at which the effective resonance frequency $\Omega_{\rmeff}$ and effective mechanical damping rate $\Gamma_ {\rmeff}$, are specified.
This equation states that the mechanical damping rate is justified by the cavity field forces.
Splitting the effect of radiation pressure and bolometric forces, the effective damping rate reads as $\Gamma_{\rmeff} \equiv \gamma_{\rmm}+\Gamma_{\rmrp}+\Gamma_{\rmpt}$, where
\begin{subequations}\label{geff}
\begin{eqnarray}
\Gamma_{\rmrp}&\equiv & \frac{2 \gamma_{\rmc} \Delta G^2 \omega_{\rmm}}{[\gamma_{\rmc}^2+(\omega_{\rmm}-\Delta)^2][\gamma_{\rmc}^2+(\omega_{\rmm}+\Delta)^2]}, \\
\Gamma_{\rmpt}&\equiv & \frac{2 \gamma_2 \beta \chi L \Delta G^{2}\omega_{\rmm} [2 \gamma_{\rmc} +\tau_{\rmth} (\Delta^2 +\gamma^2 -\omega_{\rmm}^2 )]}{(1+\tau_{\rmth}^2 \omega_{\rmm}^2) [\gamma_{\rmc}^2+(\omega_{\rmm}-\Delta)^2][\gamma_{\rmc}^2+(\omega_{\rmm}+\Delta)^2]}.
\end{eqnarray}
\end{subequations}
On the other hand, the modified mechanical frequency, which is induced by light (optical spring effect), is given by
\begin{widetext}
\begin{equation}\label{weff}
\Omega_{\rmeff} \equiv \Big\{\omega_{\rmm}^2 -\frac{\Delta G^2 \omega_{\rmm} (\gamma_{\rmc}^2 +\Delta^2 -\omega^2)}{[\gamma_{\rmc}^2+(\omega_{\rmm}-\Delta)^2][\gamma_{\rmc}^2+(\omega_{\rmm}+\Delta)^2]}
-\frac{2 \gamma_2 \beta \chi L \Delta G^{2}\omega_{\rmm} [\gamma_{\rmc}^2 +\Delta^2 -\omega^2 (1 +2\gamma_{\rmc} \tau_{\rmth})]}{(1+\tau_{\rmth}^2 \omega_{\rmm}^2) [\gamma_{\rmc}^2+(\omega_{\rmm}-\Delta)^2][\gamma_{\rmc}^2+(\omega_{\rmm}+\Delta)^2]} \Big\}^{\frac{1}{2}}.
\end{equation}
\end{widetext}
This can lead to significant change in the mechanical frequency only close to bistability threshold, but as far as the system works at weak coupling regime modification of the mechanical frequency is negligible.
It is straightforward to verify that Eqs.~(\ref{geff}) and (\ref{weff}) reproduce previous results \cite{Pinard2008,DeLiberato2011}.
Moreover, by setting $\gamma_{2}$ equal to zero, the effective frequency and effective damping rate of a pure radiation pressure optomechanical system are obtained which are already reported in Ref.~\cite{Genes2008b}.

%
%
\section{Photothermal Cooling}
In this section, we study cooling the motion of the mechanical resonator toward its ground state by means of the correlation matrix introduced in Sec.~III.
Afterwards, to have a physical intuitive picture and to compare our results to the other works, we consider a weakly coupled optomechanical system to approximate the exact results.
However, before every thing, the stability conditions must be carefully taken into account.
Because our steady state analysis is valid only for a stable system.

The system is stable when all poles of the effective susceptibility (\ref{suscept}) lie in the lower complex half plane.
The parameter region in which the system is satisfactorily stable can be obtained from the Routh--Hurwitz criteria \cite{Gradshteyn2007}.
The inequalities determining the stability conditions are too cumbersome to be reported here.
However, one can appreciate this from the characteristic polynomial of the effective susceptibility (see Appendix B).
In the case of a red detuned cavity ($\Delta>0$), which is the interesting regime both for cooling and entanglement, the only nontrivial stability condition is
\begin{equation}
\omega_{\rmm}(\gamma_{\rmc}^2 +\Delta^2) -\Delta G^{2} \big(1 +2 \gamma_2 \beta \chi L)>0.
\label{stability}
\end{equation}
In the rest of the paper satisfaction of this stability condition is considered.

The effective phonon occupation number of the mechanical resonator, $n_{\rmeff}$, can be derived from the total mechanical energy
\begin{equation}
\label{mecen}
E_{\rmm}=\frac{1}{2}\hbar \omega_{\rmm}[\langle\delta p^2 \rangle +\langle \delta q^2 \rangle] = \hbar \omega_{\rmm}(n_{\rmeff}+\frac{1}{2}).
\end{equation}
The mechanical resonator approaches to its ground state if $n_{\rmeff} \simeq 0$ or equivalently $E_{\rmm} \simeq \hbar \omega_{\rmm}/2$.
Therefore, to analyze the cooling process, one needs to calculate the mechanical position and momentum variances, which are respectively first and second diagonal elements of the CM.
We have already discussed about $\langle \delta q^2 \rangle$, while variance of the mechanical momentum is simply related to the position variance by
\begin{equation}
\label{varp}
\langle \delta p^2 \rangle = \int{\frac{\rmd \omega}{2 \pi} \frac{\omega^2}{\omega_{\rmm}^2} \mcS_{qq}(\omega) }.
\end{equation}
The integrals in (\ref{varq}) and (\ref{varp}) can be performed for a stable system.
However, a sensible approximation on the Brownian noise equation, significantly simplifies the final result.
In fact, the Brownian noise $\xi(t)$ is in general a non Markovian Gaussian noise [see Eq.~(\ref{sth})].
Nevertheless, to observe the quantum mechanical effects of the mechanical resonator, it is necessary to prepare a high quality mechanical resonator $Q_{\rmm}\gg 1$.
At this limit $\xi(t)$ becomes Markovian and we have
\begin{equation}
\frac{\gamma_{\rmm} \omega}{\omega_{\rmm}}  \coth[\frac{\hbar \omega}{2 k_{\mathrm{B}} T}] 
\simeq \gamma_{\rmm} \frac{2 k_{\mathrm{B}} T}{\hbar \omega_{\rmm}} 
\simeq \gamma_{\rmm} (2 \bar{n}+1),
\end{equation}
where $\bar{n}=[\exp(\hbar \omega_{\rmm}/k_{\mathrm{B}T})-1]^{-1}$ is mean thermal phonon number of the mechanical resonator connected to a reservoir at equilibrium temperature $T$.

Another important point must be considered here is that the ground state cooling is not approached only by attaining $n_{\rmeff} < 1$ .
In fact, in order to get the quantum ground state cooling, it is also necessary for both the mechanical displacement and momentum variances to satisfy the minimum uncertainty principle.
Strictly speaking, both variances have to tend to $1/2$, therefore, energy equipartition has to be satisfied in the optimal regime close to the ground state.
It is essential to be pointed out that, in general, energy equipartition is not necessarily true as it can be seen from (\ref{varq}) and (\ref{varp}) that the two variances are not equal  $\langle \delta q^2 \rangle \neq \langle \delta p^2 \rangle$.

\subsection{Weak coupling limit}
Performing the integrals for $\langle \delta q^2 \rangle$ and $\langle \delta p^2 \rangle$, and then replacing in (\ref{mecen}), gives an exact result for the effective mechanical phonon number, which is too involved to be reported here.
However, at weak coupling regime $\gamma_{\rmc} \gg \gamma_{\rmm},G$, one gets an approximate expression for the effective phonon number.
Additionally, the cooling process can be executed at low temperatures, at which number of the thermal phonons is very small $\omega_{\rmm} \gg \bar{n} \gamma_{\rmm}$.
Exerting the above limits; weak optomechanical regime and working at cryogenic temperatures, we arrive at the following approximate equation for the effective phonon number
\begin{equation}
n_{\rmeff}=\frac{\bar{n} \gamma_{\rmm} +A_{\rmrp}^+ +A_{\rmpt}^+ +A_{\rmcc}^+}{\gamma_{\rmm}+\Gamma_{\rmrp}+\Gamma_{\rmpt}},
\label{neff}
\end{equation}
where we have introduced the photon scattering rates $A_{\rmrp}^+$, $A_{\rmpt}^+$, and $A_{\rmcc}^+$ corresponding to the radiation pressure, photothermal force, and their cross correlation effect, receptively.
These are the rates at which the laser photons are scattered by the mechanical oscillator simultaneously with the absorption of the oscillator vibrational phonons, given by 
\begin{subequations}\label{rates}
\begin{eqnarray}
A_{\rmrp}^+ &=& \frac{G^2 \gamma_{\rmc}}{2 [\gamma_{\rmc}^2+(\omega_{\rmm}+\Delta)^2]}, \\
A_{\rmpt}^+ &=& \frac{\gamma_2 (\beta \chi L)^{2} G^{2}}{2(1+\tau_{\rmth}^2 \omega_{\rmm}^2)}, \\
A_{\rmcc}^+ &=& \frac{\gamma_2 \beta \chi L G^{2}[\gamma_{\rmc} -\tau_{\rmth}  \omega_{\rmm} (\Delta +\omega_{\rmm})]  }{(1+\tau_{\rmth}^2 \omega_{\rmm}^2) [\gamma_{\rmc}^2+(\omega_{\rmm}+\Delta)^2]}.
\end{eqnarray}
\end{subequations}
Interestingly, it is obvious from Eqs.~(\ref{rates}c) that for a red detuned cavity, depending on the working regime, $A_{\rmcc}^{+}$ can get positive or negative values.
Actually, for an optomechanical system in bad cavity regime ($\gamma_{\rmc} \gg \omega_{\rmm}$) $A_{\rmcc}^{+}$ gets positive values, while for a good cavity system ($\gamma_{\rmc} \ll \omega_{\rmm}$) it becomes negative.
Such a change in the sign of $A_{\rmcc}^{+}$ at the good cavity regime can play an important role in the final effective phonon number.
It is notable that in this regime the radiation pressure cools down the mechanical resonator very effectively \cite{Marquardt2007,Wilson-Rae2007}.

One can also obtain Eq.~(\ref{neff}) by employing the asymmetric noise spectrum method exploited for the radiation pressure cooling in Ref.~\cite{Marquardt2007}.
According to the method, the minimum attainable phonon number is given by the following equation
\begin{equation}\label{nmin}
n_{\mathrm{min}}\equiv \Big[\frac{S_{\mathrm{opt}}(-\omega_{\rmm})}{S_{\mathrm{opt}}(+\omega_{\rmm})}-1 \Big]^{-1},
\end{equation}
where $S_{\mathrm{opt}}(\omega)$ is the asymmetric noise spectrum of the optical force.
In Ref.~\cite{Restrepo2011} the noise spectrum of the light forces, including both radiation pressure and photothermal pressure, was applied to Eq.~(\ref{nmin}) to get the effective mechanical phonon occupation.
Considering the difference in definition of the elasticity constant (one needs to replace $2\beta/c$ by $\beta\chi$), coincidence of our result with the result of Ref.~\cite{Restrepo2011} can be easily verified.
Moreover, it can be shown that suppressing $A^{+}_{\rmcc}$ in Eq.~(\ref{neff}) gives the result obtained in Ref.~\cite{DeLiberato2011}.
In fact, in Ref.~\cite{DeLiberato2011} it was supposed that the radiation pressure and photothermal force operate independently, so the cross correlation effect in photothermal cooling is omitted.
Our solution is also in coincidence with the quantum theory of radiation pressure cooling \cite{Genes2008b,Wilson-Rae2007,Marquardt2007}.
Actually, if no photon is absorbed by the micromechanical mirror, i.e. $\gamma_2 =0$, Eq.~(\ref{neff}) reproduces the result obtained by Genes \emph{et al}.

\subsection{Cooling at strong coupling regime}
Now let us go beyond the previous works and study the photothermal cooling at strong coupling regime.
We use the exact effective phonon number obtained from Eq.~(\ref{mecen}) to make a numerical investigation on the cooling process.
In the following analysis, we implicitly inspect satisfaction of the energy equipartition which is, as mentioned above, an important condition for ground state cooling.
In fact, in the parameter region utilizing in this paper --both for bad and good cavities-- it is always true that $\langle \delta q^2 \rangle \simeq \langle \delta p^2 \rangle$.

\begin{figure}
\label{phonon}
\includegraphics[width=3.4 in]{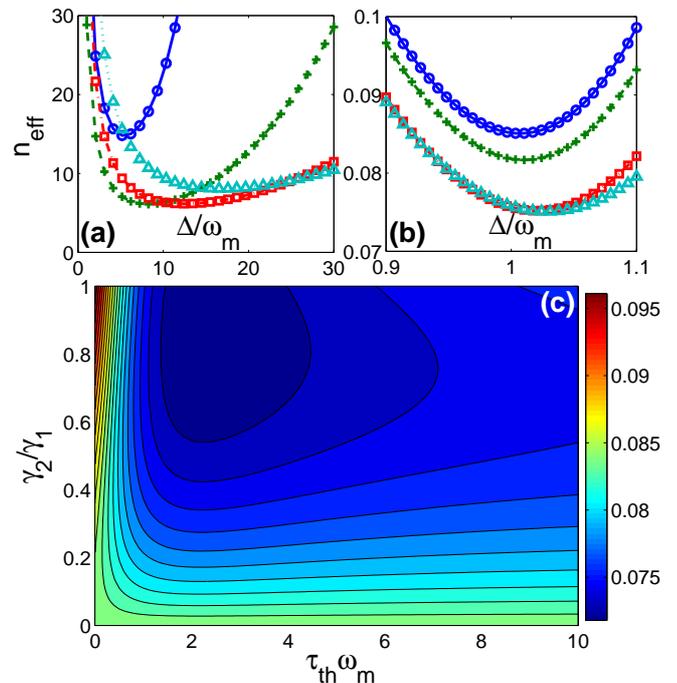}
\caption{(Color online) Variation of the effective phonon number versus normalized cavity detuning for a bad cavity $\gamma_1 /\omega_{\rmm} =10$ (a) and a good cavity $\gamma_1 /\omega_{\rmm} =0.1$ (b):
Bare radiation pressure system (blue circle) and for a system including both radiation pressure and photothermal effects; $\gamma_2 /\gamma_1 =0.1$ (green plus), $\gamma_2 /\gamma_1 =0.5$ (red square) and $\gamma_2 /\gamma_1 =1.0$ (cyan triangle).
The contour plot of inset (c) presents the effective phonon number versus normalized thermal diffusion time and normalized decay rate from the micromechanical mirror side (photon absorption rate) for a good cavity ($\gamma_{1}/\omega_{\rmm}=0.1$) at fixed cavity detuning $\Delta=\omega_{\rmm}$.
The input power is $P =15$ mW and the rest of the system parameters are same as the parameters used in Fig.~2.}
\end{figure}

In Fig.~3(a) and (b), effective phonon number of the micromechanical resonator is plotted versus cavity detuning for a bad cavity and a good cavity, respectively.
The parameters we are considering here correspond to a practically attainable setup, for example one can operate the mechanical resonator in Ref.~\cite{Verbridge2008} as the movable mirror of an optomechanical system similar to the setup in Ref.~\cite{Favero2009a}.
The figure shows that for a bad cavity the minimum achievable phonon number of a photothermal system is much smaller than a pure radiative system.
The minimum number of phonons in this case is about five phonons which is exactly the same as the result of Ref.~\cite{DeLiberato2011}.
On the other hand, surprisingly the photothermal force has a positive effect on the ground state cooling even in a good cavity setup.
This can be found out from Fig.~3(b), where the minimum phonon number accomplished by a pure radiation pressure system is about $0.086$, while adjustment of photothermal effects (by adjusting the rate of photon absorption at the micromechanical mirror) reduces it to about $0.075$.
Therefore, better cooling results can be achieved by engineering the photon absorption rate of the micromechanical mirror.
For example, one can control the absorption rate $\gamma_{2}$ by spreading gold nanoparticles on the surface of the cantilever.
Adjusting the nanoparticle size, justifies its optical absorption coefficient, which in turn modifies absorption rate of the mechanical resonator.

However, the optomechanical cooling can be optimized more by another characteristic parameter of the photothermal force, that is the thermal diffusion time $\tau_{\rmth}$, which also functions as the delay time in the feedback loop.
In Fig.~3(c) we have depicted variation of the effective phonon number with the absorption rate, $\gamma_2$, and thermal diffusion time, $\tau_{\rmth}$, for a good cavity system.
It is evident from the figure that there is an optimum value for both absorption rate and thermal diffusion time, which gives the minimum effective phonon number.
Employing these optimal values in a photothermal system, the effective phonon number can even be reduced to $0.072$.
This is equivalent to an effective temperature about $178$ $\mu$K for the mechanical mode.

It is also interesting to compare the exact result, which enables us to consider the system at strong coupling regime, with the approximate calculation.
For the optomechanical parameters employed in this paper, a bad cavity system is always working at weak coupling limit.
Therefore, the approximate result reproduces just the exact solution.
However, for a good cavity, the weak coupling limit is only obtained for small input powers.
Thus, we focus on the good cavity regime to see how can strong optomechanical coupling affect the ground state cooling.
For this purpose, in Fig.~4 we have plotted the effective mechanical phonon number versus input laser power which mostly determines value of the effective optomechanical coupling.
In the figure, the exact and approximate values of $n_{\rmeff}$ are included both for a pure radiative system and a photothermal scheme.
The dashed vertical green line indicates the laser power which gives $G=\gamma_{\rmc}$, therefore, the right hand side of the line absolutely refers to the strong coupling region.
It can be seen that for $P<2$ mW, which pertains to the weak regime, the exact and approximate results begin to coincide, while for larger values of $P$ the approximate lines go far from their corresponding exact lines.
Nevertheless, one can conclude from Fig.~4 that the ground sate cooling is enhanced at strong coupling regime, especially photothermal effects are more responsive at this regime.
Additionally, the figure shows that there is a finite optimal value for $G$ which gives the minimum $n_{\rmeff}$.
Hence, an arbitrary large $G$ does not necessarily improve the ground state cooling.

\begin{figure}
\includegraphics[width=3.1 in]{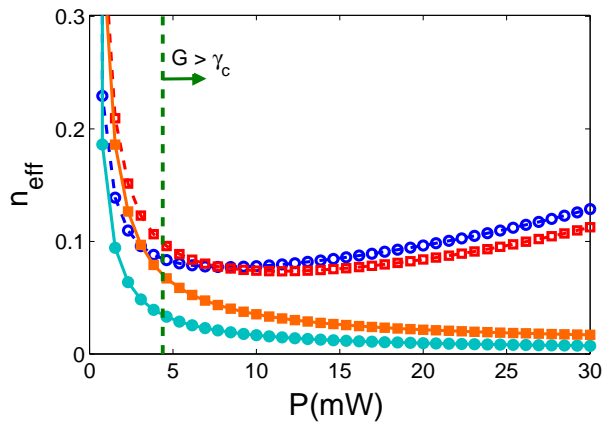}
\caption{(Color online) Variation of the effective phonon number versus laser input power for a good cavity $\gamma_1 /\omega_{\rmm} =0.1$.
For a pure radiative system; exact solution (blue circle) and approximate solution (cyan filled circles).
For a photothermal scheme with $\gamma_{2} /\gamma_{1} = 1$; exact solution (red square) and approximate solution (orange filled square).
The cavity detuning is $\Delta = \omega_{\rmm}$ and the rest of the system parameters are same as the parameters used in Fig.~2.}
\end{figure}

Finally, let us check validity of our theoretical formulation with already existing experimental evidence, and then investigate availability of better results by the same experimental setup.
According to our results, the effective temperature of the mechanical resonator is about $30$ K for $\gamma_1 /\omega_{\rmm} = 800$ and $\Delta/\omega_{\rmm}=650$.
These are nearly the same experimental parameters utilized in Ref.~\cite{Metzger2008} to achieve a cantilever cooled down to 30 K, therefore, our analysis gives a good insight for operation of the photothermal effect.
On the other hand, for the parameters of the setup in Ref.~\cite{Favero2007} (very similar to that of Ref.~\cite{Metzger2008}), we have computed effective temperature of the mechanical resonator for a wide range of cavity detuning $\Delta$ and the cavity input side decay rate $\gamma_1$.
Our numerical results indicate that, even though the authors in Ref.~\cite{Favero2007} --starting from room temperature-- achieved a minimum temperature about 70 K for the mechanical resonator, one is able to cool it down to effective temperatures as less as 0.2 K.
In fact, employing the same device, it is possible to achieve such a low temperature by improving the cavity finesse such that $\gamma_{1}/\omega_{\rmm} \approx 60$, and adjusting the laser frequency detuning.

%
%
\section{Photothermal Entanglement}
Eventually, in this section we shall discuss about the most important result of the paper; the optomechanical entanglement in a system coupled through both radiation pressure and photothermal force.
Entanglement between various constituents of a radiation pressure based optomechanical system is thoroughly studied, including optomechanical entanglement of a mechanical resonator and optical cavity modes and also all optical entanglement of the output modes (see e.g. Ref.~\cite{Genes2008}).
However, effect of photothermal pressure on the optomechanical entanglement which can be an interesting concept is not studied, so far.
The optomechanical entanglement is produced only in a strongly coupled system, therefore, only  a considerable contribution of the photothermal force can enhance it.
In fact, although the photothermal force naturally introduces decoherence, yet a strong photothermal action even can improve the optomechanical entanglement.
To investigate the situation, we first provide an exact numerical description, and then an approximate analytical analysis on the entanglement is presented.

In the following we use logarithmic negativity as a convenient measure for continuous variable entanglement.
Log negativity is defined as \cite{Vidal2002,Adesso2004}
\begin{equation}
E_N =\mathrm{max}[0,-\ln 2 \eta^-],
\label{logneg}
\end{equation}
where $\eta^-$ is the minimum symplectic eigenvalue of partial transposed CM corresponding to the bipartite system of interest, which is given by
\begin{equation}
\eta^- \equiv 2^{-1/2}\{\Sigma(V)-\sqrt{\Sigma(V)^2 -4\det V}\}^{1/2},
\label{etamin}
\end{equation}
with $\Sigma(V) \equiv \det V_{\mathrm{A}} +\det V_{\mathrm{B}} -2\det V_{\mathrm{C}}$.
Here we have considered the CM in a $2 \times 2$ block form as
\begin{equation}
V = \left(\begin{array}{cc}
	V_{\mathrm{A}} & V_{\mathrm{C}} \\
	V_{\mathrm{C}}^{\mathsf{T}} & V_{\mathrm{B}} \\
	\end{array}\right).
\end{equation}
Therefore, from (\ref{logneg}) and (\ref{etamin}) one concludes that a Gaussian state is entangled if and only if $\eta^- <1/2$, which is equivalent to Simon’s necessary and sufficient entanglement nonpositive partial transpose criterion for Gaussian states \cite{Simon2000}.

\begin{figure}
\label{entanglement}
\includegraphics[width=3.4 in]{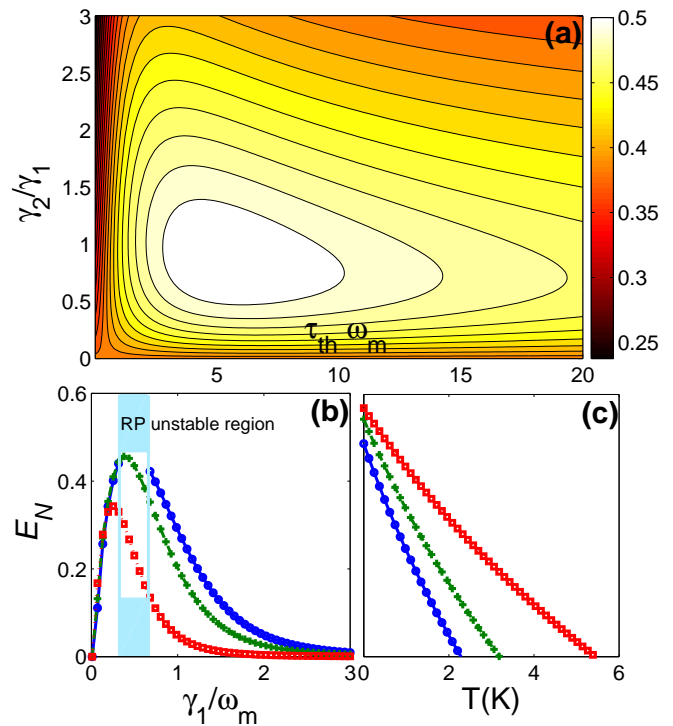}
\caption{(Color online) The contour plot of logarithmic negativity versus normalized thermal diffusion time and normalized decay rate from the micromechanical mirror side (a).
Variation of log negativity versus normalized input mirror decay rate (b). Variation of log negativity versus temperature of the mechanical reservoir (c):
In insets (b) and (c) blue circle corresponds to a bare radiation pressure system, while the other lines belong to a system including both radiation pressure and photothermal effects; $\gamma_{2}/\gamma_{1}=0.2$ (green plus), $\gamma_{2}/\gamma_{1}=0.9$ (red square).
In the case of inset (a) and (c) the input side decay rate and the laser power are $\gamma_{1}/\omega_{\rmm}=0.1$ and $P=50$ mW, respectively.
In the inset (b), the cyan region indicates the unstable region of a pure radiation pressure system and the input power is $P=20$ mW.
The cavity detuning is always fixed at $\Delta/\omega_{\rmm}=0.85$. The rest of the system parameters are same as the parameters used in Fig.~2.}
\end{figure}

\subsection{Exact numerical results}
Now let us study behaviour of entanglement between the mechanical resonator and the optical cavity mode, $E_N$, from exactly calculated CM (\ref{cm}).
We shall focus on the two exclusive photothermal parameters, the characteristic time of the thermal diffusion $\tau_{\rmth}$ and the photon absorption rate of the movable mirror $\gamma_2$, to examine their effect on the optomechanical entanglement.
We have made a careful analysis in a wide parameter range and we found a parameter region very close to performable photothermal experiments \cite{Verbridge2008,Favero2009a}.

In Fig.~5(a), variation of $E_N$ versus $\gamma_{2}$ and $\tau_{\rmth}$ is shown by a contour plot.
The lower boundary of the plot, corresponding to $\gamma_2 / \gamma_1 =0$, refers to a system which is coupled only by the radiation pressure.
It is clear from the figure that for a radiative system, the log negativity is about $0.4$, which --as expected-- is independent of $\tau_{\rmth}$.
At the first sight one can see that, interestingly, increasing $\gamma_2$, which brings in photothermal effects, can improve the entanglement.
One can see from Fig.~5(a) that for small values of the thermal diffusion time ($\tau_{\rmth} \omega_{\rmm} <1$), $E_N$ decreases rapidly with $\gamma_{2}$, therefore, it can be concluded that a fast photothermal action --compared to the mechanical response time $1/\omega_{\rmm}$-- severely destroys the entanglement.
However, at larger $\tau_{\rmth}$ values, $E_{N}$ reaches to its maximum for a photothermal system ($\gamma_2 \neq 0$).
This consequence means that in some cases, the photothermal force can lead to enhancement of the optomechanical entanglement.
Especially, in the parameter regime selected in this paper, the log negativity increases significantly to $0.5$ for an absorption rate $\gamma_{2}$ equal to the input mirror decay rate $\gamma_{1}$.

Instead, in Fig.~5(b), we plot the log negativity respect to the input decay rate $\gamma_{1}$ at fixed pump power and cavity detuning.
The teal lines indicate the parameter region at which the pure radiation pressure system is unstable.
Interestingly, from the figure it is obvious that the photothermal force can significantly modify the stability region.
In fact, the optomechanical entanglement of a pure radiative system cannot raise enough due to instability of the system, while in the presence of photothermal effects the bistability threshold retreats, therefore, $E_{N}$ can raise to its maximum.
However, the raising slope of $E_{N}$ in the photothermal case is smaller than the radiation pressure case, since the photothermal force introduces decoherence to the system.
According to the figure, further increase in $\gamma_{2}$ leads to destruction of the entanglement, as decoherence of the system grows with it.

Now let us investigate the interesting concept of the entanglement robustness against temperature of the mechanical reservoir.
To show effect of the photothermal force on the robustness, in Fig.~5(c) the log negativity of radiation pressure and two different photothermal systems is depicted versus temperature of the mechanical resonator environment.
The figure shows that in addition to the amount, introducing an absorbing micromechanical mirror to the optomechanical system can even increase robustness of $E_{N}$.
Thus, one can enhance the robustness of $E_{N}$ by controlling absorption rate of the micromechanical mirror $\gamma_{2}$, which can be achieved by spreading metallic nanoparticles on the surface of the mirror.

\subsection{Approximate analytical results}
The exact expression of the logarithmic negativity stemming from the solution of (\ref{fcm}) is very cumbersome, but it is nonetheless possible to explain some of the above results by means of an approximate treatment which satisfactorily describes the effect of photothermal force on $E_{N}$.
We derive an approximate analytical expression for $E_{N}$ in the parameter region very close to the bistability threshold.
This regime is relevant for optomechanical entanglement because, as first pointed out in Ref.~\cite{Genes2008} and later discussed in detail in Ref.~\cite{Ghobadi2011}, $E_{N}$ reaches its maximum value at the bistability threshold.
Therefore, taking the threshold value for the optomechanical coupling $G \simeq \sqrt{\omega_{\rmm}(\gamma_{\rmc}^{2}+\Delta^{2})/[\Delta(1+2\gamma_{2}\beta \chi L)]}$, and neglecting thermal noise terms one gets the following approximate expression for the minimum symplectic eigenvalue
\begin{equation}\label{hmin}
\eta^{-} \simeq \Big\{\frac{\gamma_{\rmc} \big[\omega_{\rmm}^{4}+4 \Delta^{2} (\Delta^{2}+\gamma_{\rmc}^{2}+\omega_{\rmm}^{2})\big] A_{\eta} +\beta \chi L \gamma_{2}B_{\eta}}{2\big[8 \gamma_{\rmc} \Delta^{2} (\Delta^2 +\gamma_{\rmc}^2 + 5 \omega_{\rmm}^{2}) A_{\eta} + \beta \chi L \gamma_{2}C_{\eta}\big]} \Big\}^{\frac{1}{2}},
\end{equation}
where
\begin{align}
A_{\eta} &\equiv 1+\tau_{\rmth} \big[\Delta^{2} \tau_{\rmth}+\gamma_{\rmc} (2+\gamma_{\rmc} \tau_{\rmth})+\omega_{\rmm}^{2}\tau_{\rmth} (1+2 \gamma_{\rmc}  \tau_{\rmth})\big], \nonumber \\
B_{\eta} &\equiv b_{0} +b_{2}\omega_{\rmm}^{2} +b_{4}\omega_{\rmm}^{4} +b_{6}\omega_{\rmm}^{6}, \nonumber \\
C_{\eta} &\equiv 2b_{0} +c_{2}\omega_{\rmm}^{2} +c_{4}\omega_{\rmm}^{4}, \nonumber
\end{align}
in which we have
\begin{widetext}
\begin{align}
b_{0} =& 40\Delta^{2}\gamma_{\rmc}(\Delta^{2}+\gamma_{\rmc}^{2})(1+2\gamma_{\rmc}\tau_{\rmth}+\Delta^2 \tau_{\rmth}^2+\gamma_{\rmc}^2 \tau_{\rmth}^2), \nonumber \\
b_{2} =& 8\Delta^2 \big[6 \gamma_{\rmc}^4 \tau_{\rmth}^3 -\Delta^2 \tau_{\rmth} +\gamma_{\rmc}^5 \tau_{\rmth}^4+\gamma_{\rmc}^3 \tau_{\rmth}^2 (3+2 \Delta ^2 \tau_{\rmth}^2) +\gamma_{\rmc}^2 \tau_{\rmth} (7+6 \Delta^2 \tau_{\rmth}^2)+\gamma_{\rmc}(5+7 \Delta^2 \tau_{\rmth}^2+\Delta^4 \tau_{\rmth}^4)\big], \nonumber \\
b_{4} =& 2\big[\gamma_{\rmc}^3 \tau_{\rmth}^2 (3+4 \Delta^2 \tau_{\rmth}^2)+8 \gamma_{\rmc}^2 (\tau_{\rmth}+3 \Delta^2 \tau_{\rmth}^3) +\gamma_{\rmc} (5+7 \Delta^2 \tau_{\rmth}^2+4 \Delta^4 \tau_{\rmth}^4)-6 \Delta^2 \tau_{\rmth}\big], \nonumber \\
b_{6} =& 2 \tau_{\rmth} \big[6 \gamma_{\rmc}^2 \tau_{\rmth}^2+\gamma_{\rmc}^3 \tau_{\rmth}^3+\gamma_{\rmc} \tau_{\rmth}(\Delta^2 \tau_{\rmth}^2 -1)-2\big], \nonumber \\
c_{2} =& 16 \Delta^2 \big[10 \gamma_{\rmc}^4 \tau_{\rmth}^3+\gamma_{\rmc}^5 \tau_{\rmth}^4+\gamma_{\rmc}^3 \tau_{\rmth}^2 (27+2 \Delta^2 \tau_{\rmth}^2) +\Delta^2 \tau_{\rmth} (3+4 \Delta^2 \tau_{\rmth}^2)+\gamma_{\rmc}^2 \tau_{\rmth} (43+14 \Delta^2 \tau_{\rmth}^2) \nonumber \\
&+\gamma_{\rmc}  (21+31 \Delta^2 \tau_{\rmth}^2+\Delta^4 \tau_{\rmth}^4)\big], \nonumber \\
c_{4} =& 16 \Delta^2 \tau_{\rmth} \big[2 \Delta^2 \tau_{\rmth}^2+24 \gamma_{\rmc}^2 \tau_{\rmth}^2+5 \gamma_{\rmc}^3 \tau_{\rmth}^3 +\gamma_{\rmc} \tau_{\rmth} (5 \Delta^2 \tau_{\rmth}^2-3)-7\big]. \nonumber
\end{align}
\end{widetext}

If the photothermal effect is negligible, $\gamma_{2}=0$, one gets
\begin{equation}
\eta^{-} \approx \Big[\frac{\omega_{\rmm}^{4}+4 \Delta^{2} (\gamma_{\rmc}^{2}+\Delta^{2}+\omega_{\rmm}^{2})}{16 \Delta^{2} (\Delta^2 +\gamma_{\rmc}^2 + 5 \omega_{\rmm}^{2})\big]} \Big]^{\frac{1}{2}},
\end{equation}
which is in coincidence with the results of Refs.~\cite{Abdi2011,Ghobadi2011}.
Instead, in the presence of the photothermal force the inequality $\gamma_{2}<\gamma_{\rmc}$ imposes that the terms proportional to $\gamma_{2}$ in Eq.~(\ref{hmin}) cannot make a predominant contribution either in the numerator and denominator.
Hence, according to the parameter set employed in this paper, the photothermal force can slightly modify the optomechanical entanglement produced by the radiation pressure.
Such a modification in some cases can lead to enhancement of the entanglement.
In fact, $b_{0}$ is the prevailing factor both in $B_{\eta}$ and $C_{\eta}$, while in the latter it is multiplied by two.
Therefore, in general it is true that $B_{\eta}<C_{\eta}$, which means that the value of $\eta^{-}$ becomes less (so $E_{N}$ becomes greater) than its value for a pure radiation pressure system.
However, depending on the system parameters it is also possible to have $B_{\eta} > C_{\eta}$, and therefore $\eta^{-}$ may easily tend to 1/2 (so $E_{N}$ tends to zero), even for not too large values of the absorption rate $\gamma_{2}$.
Nevertheless, the other effect of the photothermal force, as discussed above, is adjustment of the bistability threshold which allows for raising of the $E_{N}$, meanwhile a pure radiative system is unstable.
Thus, the entanglement in an photothermal system can be higher than a mere radiation pressure scheme even if $B_{\eta} > C_{\eta}$.

%
%
\section{Conclusion}
We have studied the entanglement of a micromechanical mirror with a cavity mode and cooling of the resonator in the presence of photothermal effects.
We have analyzed the dynamics by adopting a standard quantum Langevin treatment.
Dynamics of the quantum fluctuations which is expressed by the linearized QLEs around the classical stationary state of the system have been analyzed.
We have investigated dependence of the log negativity and the mechanical occupancy of the optomechanical stationary state upon various system parameters.
We have also derived approximate, but compact, analytical expressions for effective mechanical phonon number in a photothermal system at weak coupling regime and for the minimum symplectic eigenvalue of the bipartite system (composed of mechanical resonator and cavity mode) close to the bistability threshold.
At the strong coupling regime we have shown numerically that the cooling is improved by the photothermal coupling.
Finally, it is realized from a numerical study and verified analytically that even though bolometric force is naturally a dissipative effect, it can improve the optomechanical entanglement.

%
%
\section*{Acknowledgments}
M.A. is grateful to D.~Vitali for useful discussions.
The authors would like to thank H.~R.~Afshar for his comments on the manuscript.

\appendix

%
%
\section{}
The thermoelastic wave equation is given by
\begin{align}
\varrho \ddot{\bfu} =& \frac{E_{\rmY}}{2(1+\sigma)(1-2\sigma)}\nabla(\nabla.\bfu) +\frac{E_{\rmY}}{2(1+\sigma)}\nabla^2 \bfu \nonumber \\
&- \frac{E_{\rmY} \alpha_{\rmth}}{3(1-2\sigma)} \nabla T,
\end{align}
where $\bfu$ is the displacement vector, $E_{\rmY}$ is Young modulus and $\sigma$ is Poisson ratio of the material constructing the mirror \cite{Landau1970}.
Splitting the displacement vector into transversal and longitudinal components, that is $\bfu =\bfu_l +\bfu_t$ where $\nabla.\bfu_t =0$ and $\nabla \times \bfu_l =0$, the wave equation for every component reads
\begin{subequations}
\label{wave}
\begin{eqnarray}
&&\ddot{\bfu}_l -c_l^2 \nabla^2 \bfu_l = -\frac{E_{\rmY} \alpha_{\rmth}}{3 \varrho(1-2\sigma)} \nabla T, \\
&&\ddot{\bfu}_t -c_t^2 \nabla^2 \bfu_t = 0,
\end{eqnarray}
\end{subequations}
where
\begin{subequations}
\begin{eqnarray}
c_l &\equiv& [\frac{E_{\rmY}(1-\sigma)}{\varrho(1+\sigma)(1-2\sigma)}]^{1/2}, \\
c_t &\equiv& [\frac{E_{\rmY}}{2\varrho(1+\sigma)}]^{1/2},
\end{eqnarray}
\end{subequations}
are speed of longitudinal and transverse thermoelastic wave propagation, respectively.

%
%
\section{}
Characteristic polynomial of the effective susceptibility Eq.(\ref{suscept}) is $P(\varphi)=\varphi^5 +p_1 \varphi^4 +p_2 \varphi^3 +p_3 \varphi^2 +p_4 \varphi +p_5$, where
\begin{subequations}
\begin{eqnarray}
p_1 &=&\gamma_{\rmm} +2\gamma_{\rmc} +\frac{1}{\tau_{\rmth}}, \\
p_2 &=&\Delta^2 + \omega_{\rmm}^2 +\gamma_{\rmc}^2 +2\gamma_{\rmc}\gamma_{\rmm} +\frac{\gamma_{\rmm} +2\gamma_{\rmc}}{\tau_{\rmth}}, \\
p_3 &=&\frac{1}{\tau_{\rmth}} \big\{ \Delta^2 (1 +\gamma_{\rmm} \tau_{\rmth}) +\omega_{\rmm}^2 \nonumber \\
&&+\gamma_{\rmc} [\gamma_{\rmc} +\gamma_{\rmm} (2 +\gamma_{\rmc} \tau_{\rmth}) +2 \tau_{\rmth} \omega_{\rmm}^2] \big\}, \\
p_4 &=&\frac{1}{\tau_{\rmth}} \big\{ (\gamma_{\rmm} +\tau_{\rmth} \omega_{\rmm}^2) (\Delta^2 +\gamma_{\rmc}^2) \nonumber \\
&&+\omega_{\rmm} (2\gamma_{\rmc}\omega_{\rmm} -G^2 \Delta \tau_{\rmth}) \big\}, \\
p_5 &=&\frac{1}{\tau_{\rmth}} \big\{ \omega_{\rmm}^2 (\Delta^2 +\gamma_{\rmc}^2) 
-G^{2} \Delta \omega_{\rmm} (1 +2 \gamma_2 \chi \beta L) \big\}.
\end{eqnarray}
\end{subequations}
%

\bibliography{photothermal}
\end{document}